\documentclass{article}
\usepackage{spconf,amsmath,graphicx}

\usepackage{amssymb}
\usepackage[utf8]{inputenc}
\usepackage{url}
\usepackage{subfig}
\usepackage{array}
\usepackage{float}
\usepackage{grffile}
\usepackage{rotating}
\usepackage{multirow}
\usepackage[dvipsnames]{xcolor}

\title{An Asymmetric Cycle-Consistency Loss for Dealing with Many-to-One Mappings in Image Translation: A Study on Thigh MR Scans}
\name{M. Gadermayr$^{\star}$ $^{\dagger}$ \quad M. Tschuchnig$^{\star}$ \quad L. Gupta$^{\dagger}$\quad N. Kr\"amer$^{\mathsection}$ \quad D. Truhn$^{\mathsection}$ \quad D. Merhof$^{\dagger}$ \quad B. Gess$^{\ddagger}$
}	

\address{$^{\star}$ Salzburg University of Applied Sciences, 5412, Austria \\
	$^{\dagger}$ Institute of Imaging \& Computer Vision, RWTH Aachen University, Aachen, Germany\\
	$^{\ddagger}$ Department of Neurology, University Hospital Aachen, RWTH Aachen University, Aachen, Germany\\
    $^{\mathsection}$ Department of Radiology, RWTH Aachen, University Hospital Aachen, 52074 Aachen, Germany}
\begin{document}
%
\maketitle
\begin{abstract}
Generative adversarial networks using a cycle-consistency loss facilitate unpaired training of image-translation models and thereby exhibit a very high potential in manifold medical applications. However, the fact that images in one domain potentially map to more than one image in another domain (e.g. in case of pathological changes) exhibits a major challenge for training the networks. In this work, we offer a solution to improve the training process in case of many-to-one mappings by modifying the cycle-consistency loss. We show formally and empirically that the proposed method improves the performance without radically changing the architecture and increasing the model complexity. We evaluate our method on thigh MRI scans with the final goal of segmenting the muscle in fat-infiltrated patients' data.
\end{abstract}
%

\vspace{-0.1cm}
\section{Motivation}\label{sec:introduction}
\begin{figure*}[hbt] \center
	\includegraphics[width=1\linewidth]{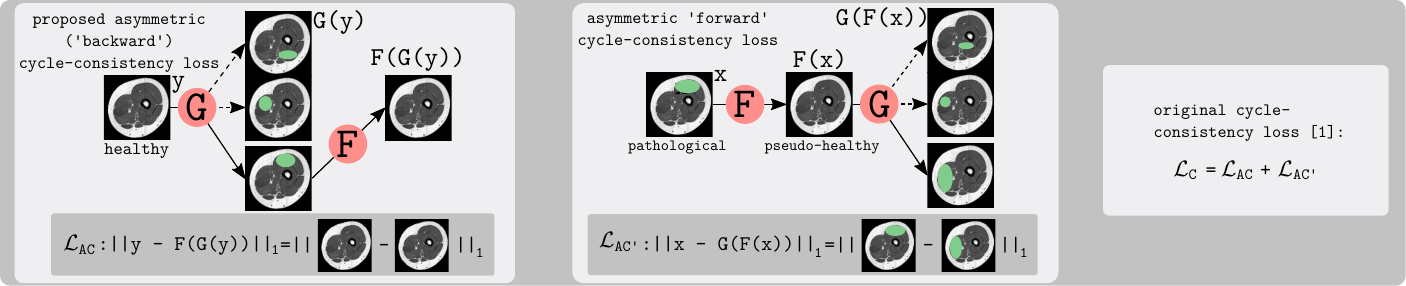}
	\caption{Schematic outline of the problem of one-to-many and many-to-one mappings with the formulation proposed in~\cite{myZhu17a}. 
		The cycle on the left side generates a small loss in case of well performing generators $F$ and $G$. 
		The cycle on the right side does not necessarily produce a small loss if $F(x)$ reliably removes all information regarding the characteristics of domain $X$.}
	\label{fig:scheme}
\end{figure*}

\vspace{-0.1cm}
The development of generative adversarial networks using the cycle-consistency loss~\cite{myZhu17a,myYi17a} 
facilitates learning image-translation between two domains without the need for any paired samples.
Over the last three years, numerous similar approaches were effectively applied in the field of medical image analysis~\cite{myWolterink17a,myJiang18a,myBentaieb18a}.
Although the methods have a very high potential in medical image computing, there are technical limitations which restrict the applicability in certain scenarios.
One major challenge is the so-called one-to-many mapping problem~\cite{myZhu17b}. If a sample of one domain can have different mapping options to another domain, the cycle-consistency loss does not provide an optimum metric for comparing a source and a reconstructed image. Even though, both images show semantically similar samples, the loss might be high due to the different mapping options (see also Fig.~\ref{fig:scheme}).
Several approaches have so far been proposed to tackle the one-to-many mapping problem in unpaired image-translation settings:
Almahairi et al.~\cite{myAlmahairi18a} and Huang et al.~\cite{myHuang18a} proposed GAN architectures employing auxiliary latent spaces to control the variations of the one-to-many (or even many-to-many~\cite{myAlmahairi18a}) mappings. The idea is to decompose an image into a content code that is domain-invariant and a style code which captures domain-specific properties.
These approaches focus on translating one single sample into a set of samples of the opposing domain.
Park et al.~\cite{myPark20a} use a patchwise contrastive loss exploiting the fact that a patch appears closer to its corresponding input than to other random patches. This enables one sided translation without assuming a bijection between the domains.
In this work, we are particularly interested in the translation direction from 'many' to 'one', which is not ambiguous. An input image should be translated to exactly one output image and we are not interested in generating several corresponding output images. Nevertheless, similar challenges during training the cycle-GAN~\cite{myZhu17a} arise since both (one-to-many \& the many-to-one) mappings need to be learned simultaneously.
Another approach to deal with many-to-one mappings relies on a separation into sub-groups~\cite{myGao19a}. While sub-groups are often naturally given (e.g. based on the scanning device or lab), in the considered scenario, transitions are flowing and a categorization is thereby completely impossible.

In this paper, we show why the original cycle-GAN approach~\cite{myZhu17a} is not optimal here and propose a modification to facilitate training without changing the general GAN architecture. 
We formally and empirically show that our approach improves image translation without increasing model complexity or setup parameters. 
In the experimental study (Sect.~\ref{sec:study}), we investigate the applicability to MRI scans of human thighs. 
We consider the generation of pseudo-healthy samples (which are easier to segment than the original unhealthy images) with the final goal of muscle segmentation~\cite{Gadermayr18a,myYao17a} to determine the fat-fraction.

\vspace{-0.1cm}
\section{Learning Many-to-One Mappings}\label{sec:methods}
\vspace{-0.1cm}
Given two domains $X$ and $Y$, we assume there is a correspondence between the elements of these domains. We assume that for each object of $X$ there is exactly one corresponding object in $Y$ (Fig.~\ref{fig:scheme}, right) and each object in $Y$ has at least one corresponding object $X$ (Fig.~\ref{fig:scheme}, left).
Formally expressed, there is a surjective function $f$, mapping objects from the domain $X$ to the corresponding objects in domain $Y$. As the function is surjective, there is also a semi-inverse $g$ such that $f \circ g$ represents the identity mapping.
Regarding the opposite direction, we assume that for each object of domain $Y$, there are several corresponding objects in domain $X$.
Consequently, $f$ is not injective (and thereby not bijective) and there is no left-inverse function $g'$ such that $g' \circ f$ represents the identity mapping.

In the original cycle-GAN approach, two generators $F$ and $G$ and two discriminators $D_X$ and $D_Y$ are optimized based on a combination of the GAN loss
and the cycle-consistency loss
\vspace{-0.1cm}
\begin{equation}
\begin{split}
\mathcal{L}_{C}=\mathbb{E}_{y \sim p_{data}(y)} [|| F(G({y})) - {y} ||_1 ] + \\
\mathbb{E}_{x \sim p_{data}(x)} [ ||G(F({x})) - {x}||_1 ] \; .
\end{split}
\end{equation}
\vspace{-0.1cm}
While the GAN loss $\mathcal{L}_{GAN}$ guides optimization to produce realistic fake images, the task of the cycle-consistency loss $\mathcal{L}_C$ is to keep the semantic structures in the images. 
By learning the identity mapping, the generators are forced to maintain the information needed for reconstruction. 
Even though reversible semantic changes could theoretically occur, it has been shown that this is unlikely~\cite{myZhu17a,myAlmahairi18a}. 
Obviously, it is easier for the network to keep the semantic structure than to learn a transformation and the corresponding inverse.




In the formulation of the cycle-consistency loss $\mathcal{L}_{C}$, it is implicitly assumed that $F$ and $G$ approximate functions $f$ and $g$ such that $g \circ f (x)$ is similar to $x$ and $f \circ g (y)$ is similar to $y$ for each element from the domains $X$ and $Y$, respectively. However, this only applies if two bijective functions $f$ and $g$ exist.
This is certainly not the case in the considered many-to-one mapping scenario.
In this case, the cycle-consistency loss forces $F$ to leave some (unwanted) originally present remnants in $F(x)$ to facilitate a reconstruction ($G(F(x))$).
But as the function $f$ is surjective, there is a semi-inverse function $g$ such that $y = f \circ g(y)$ (Fig.~\ref{fig:scheme}, left).
Accordingly, we adjust the cycle-consistency loss to
\begin{equation}
\mathcal{L}_{AC}=2 \cdot  w_c \cdot \mathbb{E}_{y \sim p_{data}(y)} [|| F(G({y})) - {y} ||_1 ] \; ,
\end{equation}
where the factor 2 is introduced to keep the ratio between the cycle-consistency loss and the GAN loss.
The multiplicative factor $w_c$ is added to facilitate an adjustment of the weight of the cycle-consistency loss (for evaluation, it is also added to the original formulation in Eq. (2)).
In~\cite{myZhu17a}, the authors noted that both directions are important to prevent training instability and mode collapse.
We agree that this might apply for settings where (almost) bijective mappings between the domains exist.
But for the setting considered here, the new formulation can lead to especially enhanced fake-$Y$ domain images because $F$ no longer needs to encode information for reconstruction in the generated images.
Furthermore, in their experiments~\cite{myZhu17a}, the weight was not adjusted (doubled) accordingly with the consequence that the cycle-consistency loss was decreased which might have caused a degradation in performance.
A visualization of the many-to-one mapping problem in the considered application is provided in Fig.~\ref{fig:scheme}.

\begin{figure}[bt]\center
	\includegraphics[width=\linewidth]{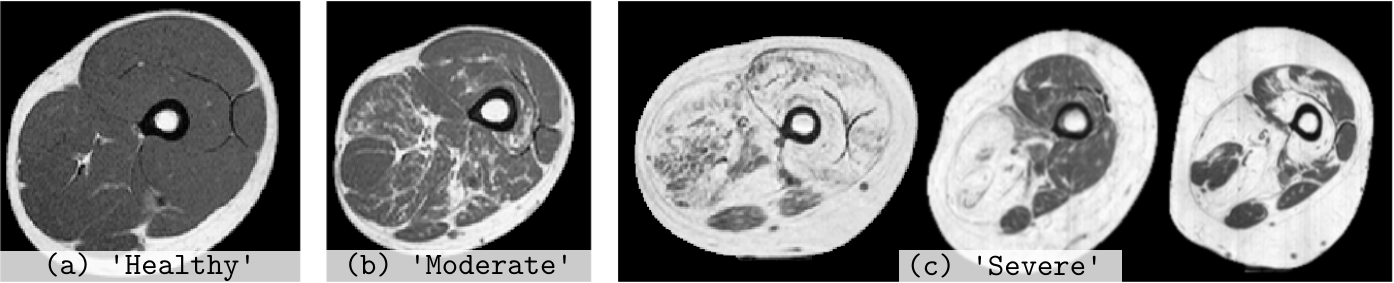}
	\caption{
		T1-weighted MRI slices in axial orientation at the thigh level showing~(a) healthy muscle, (b) moderate infiltrations and (c) largely affected muscle areas.}
	\label{fig:examples}
\end{figure}

\vspace{-0.1cm}
\section{Experimental Study}\label{sec:study}
\vspace{-0.1cm}
Neuromuscular diseases are a class of diseases caused by a variety of reasons, including gene mutations and inflammatory damage that impair the functioning of muscles either directly or indirectly through malfunctioning nerves.
They are characterized by variable degrees of fat-infiltration (Fig.~\ref{fig:examples} (b)--(c)). 
A relevant disease marker is given by the so-called fat-fraction which captures the ratio between fat and original muscle tissue within the muscular compartment. 
For computing the fat fraction, it is crucial to segment the overall muscle tissue including fat-infiltration.
Here we translate images exhibiting fat-infiltrations to pseudo-healthy images, which show no fat-infiltrations.
The rationale behind this approach is that healthy images can be segmented well based on intensity only. 
Thresholding can be used to separate muscle from other tissue which does not hold true for pathological tissue where infiltrated muscle can show similar intensity values as subcutaneous fat~\cite{Gadermayr18a}.
Fat-infiltration differs in position, size and morphology between patients. 
Due to the many-to-one mapping from diseased to healthy, we define $Y$ as the healthy domain and $X$ as the domain showing fat-infiltrations.
Instead of combining the approach with a specific segmentation approach (potentially leading to bias due to peculiarities of the chosen segmentation method), we assessed how well the images' intensity values could be reconstructed and thereby facilitate a pixel-based separation of tissue. For each image, we individually optimized an upper and a lower threshold to obtain a maximum in Dice Similarity Coefficient (DSC). This procedure is also motivated by the fact that unsupervised semi-automatic segmentation can be performed in a similar way.
Additionally, image-translation allows to extract quantitative markers related to the fat-fraction by computing the difference between the pathological images and the pseudo-healthy images in a fully-unsupervised way.
All experiments were repeated eight times. Finally the scores per patient were aggregated for each of the data sets.
In order to assess whether performances differed significantly, we applied a paired Wilcoxon signed-rank test.

\noindent
\textit{Imaging Details:}
The T1-weighted MR images were acquired on a 1.5 Tesla Phillips device with fixed echo time (17 ms), bandwidth (64 kHz) and echo train length (6) and a repetition time TR between 721 and 901 ms.  The sampling interval was fixed to 1 mm in x-y-direction and 7 mm in z-direction.
The image data consists of 21 'Healthy' and 20 'Pathological' MRI Scans showing clear fat-infiltration. 
To keep the sample size high, 2D slices extracted from the MRI scans were processed. Overall, we obtained 1124 'Healthy' and 1070 'Pathological' 2D images with a size of $256 \times 256$ pixels.
For training the image-translation models, the two domains are represented by the data sets 'Healthy' and 'Pathological'. For evaluation, the 'Pathological' data set is further divided into seven moderately ('Moderate') and 13 severely ('Severe') affected subjects as suggested in previous work~\cite{Gadermayr18a,myYao17a} (Fig.~\ref{fig:examples}).
As the category 'Healthy' can be easily segmented with existing approaches, it is not considered during evaluation (but is needed for training the translation model).
For evaluation, each fourth z-slice was annotated manually by a student assistant supervised by a medical expert. 
This strategy was applied because of a high correlation between adjacent slices and to keep the manual effort feasible.

\noindent
\textit{Network Parameters:}
Training was performed for 16 epochs with a batch-size of one and Adam as optimizer. The initial learning rate was set to $0.0001$. For data augmentation, rotation (90, 180, 270$^{\circ}$) and flipping was applied.
We applied the proposed standard cycle-GAN~\cite{myZhu17a} architecture with a U-Net~\cite{myRonneberger15a} as generator.
For both, cycle-GAN with the $\mathcal{L}_{C}$ formulation and with the $\mathcal{L}_{AC}$ approach, we investigated three settings with respect to the weighting of the losses:
We consider equal weighting of $\mathcal{L}_{GAN}$ and $\mathcal{L}_C$/$\mathcal{L}_{AC}$ ($w_c=1$) as well as a 
halved ($w_c=0.5$) and a quartered weight ($w_c=0.25$) for $\mathcal{L}_C$/$\mathcal{L}_{AC}$ to assess the impact on both approaches.
The weight of the identity loss was fixed to half the weight of $\mathcal{L}_{GAN}$.
To avoid any bias, 5-fold cross validation was performed where the folds were assigned on patient level.

\vspace{-0.1cm}
\section{Results}
\vspace{-0.1cm}

\definecolor{green}{rgb}{0.0,0.6,0.0}
\definecolor{orange}{rgb}{0.5,0.6,0.0}
\definecolor{red}{rgb}{0.8,0.0,0.0}
\renewcommand{\tabcolsep}{1.0pt}
\begin{table}[tb] \center \footnotesize
	\caption{Results obtained with (a) the asymmetric cycle-consistency loss compared to (b) the original formulation~\cite{myZhu17a} and (c) the performance with original data without any image translation. The P-value indicates whether improvements using the asymmetric loss are significant (compared to (b)).}
	\begin{tabular}{|c|rl|c|rl|rl|c|} \hline
		\multirow{2}{*}{Data Set} & \multicolumn{2}{c|}{\textbf{(c) Processing}}                    & \multicolumn{5}{c|}{Image Translation-Based Techniques} & p-value         \\ \cline{4-8}
		& \multicolumn{2}{c|}{\textbf{Original Data}}                    & $w_c$    & \multicolumn{2}{c|}{\textbf{(b) $\mathcal{L}_{C}$~\cite{myZhu17a}}} & \multicolumn{2}{c|}{\textbf{(a) $\mathcal{L}_{AC}$}} & (a) vs. (b) \\ \cline{4-8} \hline
		\multirow{3}{*}{'Moderate'} & \multirow{3}{*}{0.848} & \multirow{3}{*}{$\pm$ 0.060} & ${0.25}$  & 0.851       & $\pm$ 0.051      & 0.852        & $\pm$ 0.053        & \textcolor{red}{$>0.05$}  \\
		&                        &                              & ${0.5}$   & 0.854       & $\pm$ 0.053      & 0.859        & $\pm$ 0.046        & \textcolor{red}{$>0.05$}  \\
		&                        &                              & ${1}$     & 0.853       & $\pm$ 0.055      & 0.862        & $\pm$ 0.043        & \textcolor{red}{$>0.05$}  \\ \hline
		\multirow{3}{*}{'Severe'}     & \multirow{3}{*}{0.750} & \multirow{3}{*}{$\pm$ 0.064} & ${0.25}$  & 0.790       & $\pm$ 0.042      & \textbf{0.802}        & $\pm$ 0.047        & \textcolor{orange}{0.0391}  \\  
		&                        &                              & ${0.5}$   & 0.788       & $\pm$ 0.048      & \textbf{0.808}        & $\pm$ 0.045        & \textcolor{green}{0.0039}  \\ 
		&                        &                              & ${1}$     & 0.784       & $\pm$ 0.039      & \textbf{0.807}        & $\pm$ 0.045        & \textcolor{green}{0.0078}  \\ \hline
		\multirow{2}{*}{'Pathol.'} & \multirow{3}{*}{0.795} & \multirow{3}{*}{$\pm$ 0.081} & ${0.25}$  & 0.819       & $\pm$ 0.057      & 0.824        & $\pm$ 0.056        & \textcolor{red}{$>0.05$}  \\ 
		&                        &                              & ${0.5}$   & 0.819       & $\pm$ 0.060      & \textbf{0.831}        & $\pm$ 0.053        & \textcolor{green}{0.0067}  \\ (merged)
		&                        &                              & ${1}$     & 0.815       & $\pm$ 0.059      & \textbf{0.832}       & $\pm$ 0.053        & \textcolor{green}{0.0034} \\
		\hline
	\end{tabular}
	\label{tab:res}
\end{table}

Table~\ref{tab:res} shows quantitative results after segmentation based on thresholding individually for the 'Moderate', the 'Severe' and the combination of these two data sets ('Pathological').
Apart from the proposed $\mathcal{L}_{AC}$ loss~(a), this table also provides scores of image-translation using the original formulation ($\mathcal{L}_{C}$)~\cite{myZhu17a}~(b) and the segmentation method without image-translation~(c). 
P-values indicate whether improvements in the $\mathcal{L}_{AC}$ settings are significant (individually for each weight setting) compared to the $\mathcal{L}_{C}$ setting.
With the proposed technique, we notice increased DSCs compared to the $\mathcal{L}_{C}$ configuration (and also compared to processing without image-translation) in each configuration and each data set.
Improvements are statistically significant in five cases ($p<0.01$ in four cases and $0.01<p<0.05$ in one case). 
Generally, improvements are more distinct in case of the 'Severe' data set compared to the 'Moderate' data set. 
\begin{figure}[tb]\center
	\includegraphics[width=\linewidth]{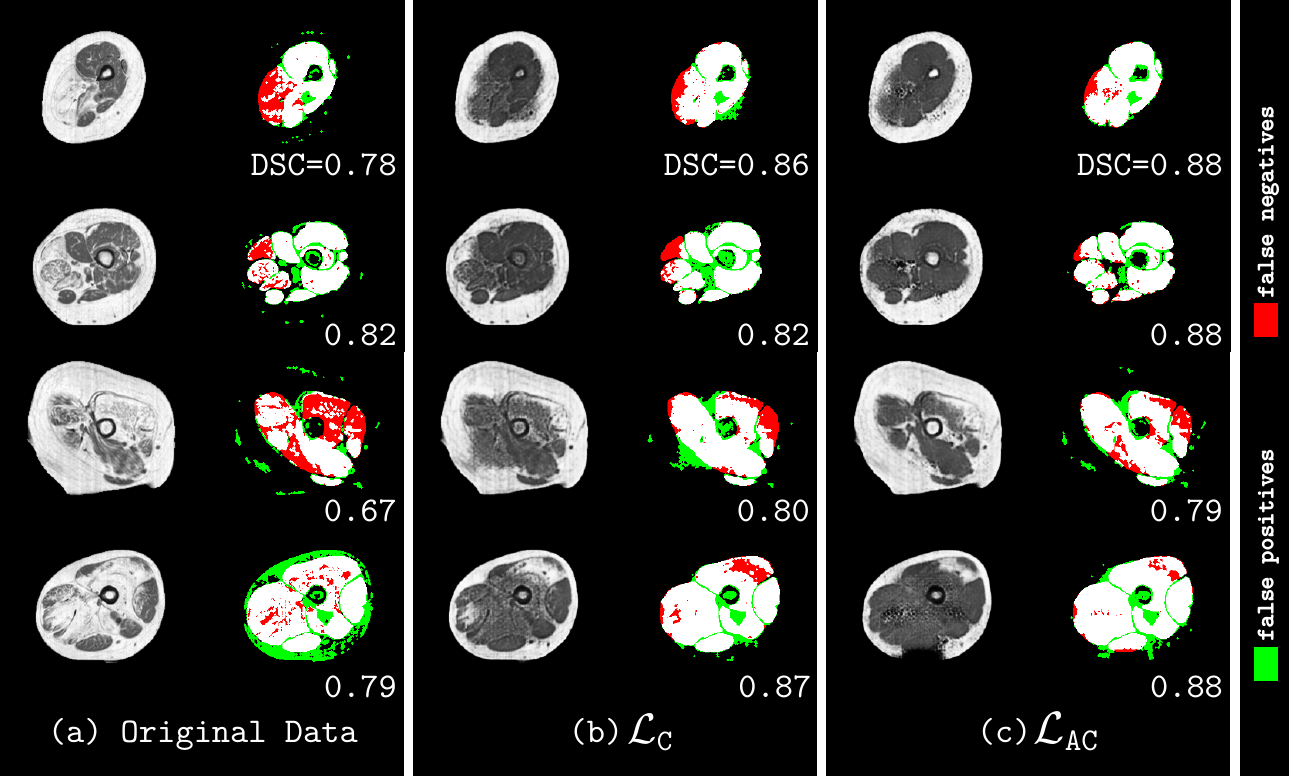}
	\caption{The left column (a) shows example original images of the 'Severe' data set including a visualization of the optimum pixel-based separability of muscle tissue using two intensity thresholds. Columns (b) and (c) show the improvements obtained with the $\mathcal{L}_{C}$ loss~\cite{myZhu17a} and with the proposed $\mathcal{L}_{AC}$ formulation.}
	\label{fig:exampleRes}
\end{figure}

The $\mathcal{L}_{AC}$ setting improved the DSCs (compared to $\mathcal{L}_{C}$) in 79 \% of 'Severe' and 62 \% of 'Moderate' 2D slices.
Considering the 'Severe' slices, DSC increases of more than 0.05 were achieved in 15 \% whereas decreases of more than 0.05 were obtained in 5 \% of the slices. 
Compared to processing of original data, we obtained improvements in 96 \% of the 'Severe' and 67 \% of the 'Moderate' 2D slices.
Fig.~\ref{fig:exampleRes} shows difficult exemplar images including corresponding visualizations of the optimum pixel-based segmentation (based on two thresholds) including the obtained DSCs for the three settings (a)--(c) presented in Table~\ref{tab:res}. 

\vspace{-0.1cm}
\section{Discussion}
\vspace{-0.1cm}
We propose a specific loss configuration facilitating the training of a cycle-GAN facing the many-to-one mapping problem.
Instead of radically changing the architecture, we showed that a slight modification of the cycle-consistency loss is capable of improving the performance without increasing the model complexity.

Based on the experimental evaluation, we observe improvements especially in case of the 'Severe' data set. Obviously, for the 'Moderate' data set, the standard approach ($\mathcal{L}_{C}$) and even a processing without any image-translation already works very well as indicated by average DSCs between 0.84 and 0.87 for all methods.
%
Considering the 'Severe' data set, we observed average improvements between 0.01 and 0.03. 
Even though the improvements are numerically decent, regarding the individual images, we notice consistent and stable improvements for the majority of cases. Stable improvements were also indicated by the positive outcomes of the significance tests (even though the number of patients was limited).
Although the performances are stable with respect to the weight configurations (the differences are not significant), we notice a trend that the $\mathcal{L}_{AC}$ configuration performs better with increasing $w_c$ compared to $\mathcal{L}_{C}$. 
We assume that this is due to the fact that our formulation is more appropriate in the specific setting, whereby overweighting of the cycle-consistency loss does not affect the outcome.

In general, we notice that the generated images do not always look perfectly realistic. We observe that image-translation techniques partly translate the subcutaneous fat into muscle tissue which finally leads to false positive segmentation areas.
This effect is expected to be due to the partial volume effect which can generate intensities similar to the intensities of muscle tissue in border regions. As this effect also occurs in healthy patients' MR scans, these regions cannot be detected as inaccurate translations by the discriminator. To prevent this issue, the data sets would need to be preprocessed and/or cleaned. 
In the generated images, we also notice artifacts which are probably due to the limited amount of training data. Even though the number of 2D slices is relatively high, the (consecutive) slices of a patient are strongly correlated. Therefore, we expect clearly improved overall quality in case of larger training data sets.
%

Comparing the qualitative output of the $\mathcal{L}_{C}$ and the $\mathcal{L}_{AC}$ setting, we notice that larger fat-infiltrations are 'filled-in' more effectively by the latter.
This is because in the original formulation, the generator $F$ needs to encode information of the location and characteristics of fat-infiltrations in the pseudo-healthy images during GAN training. Otherwise, the generator $G$ would have no chance to determine how and where to insert the fat-infiltration in order to obtain a low cycle-consistency loss. An improper placement would automatically lead to a high cycle-consistency loss as the pixel-wise $L_1$ norm is computed. In the proposed setting, this cycle is omitted and only the cycle starting with a healthy image is considered. The generator $G$ can generate fat-infiltration at any position with any realistic characteristics. As long as $F$ can recover the original muscle tissue, a low cycle-consistency loss can be obtained when the generators work effectively.



Compared to related work~\cite{myAlmahairi18a,myHuang18a}, we showed that a clearly more straightforward approach with less (hyper) parameters is similarly effective to tackle the many-to-one mapping problem. In addition, the methods proposed in~\cite{myAlmahairi18a,myHuang18a} were optimized for many-to-one mappings in which the ambiguity (e.g. length/color of hair, color of shoes) can be represented by a very limited number of parameters in a latent space. In the considered application scenario, this is not the case. The pathological modifications considered in this paper can show arbitrary shape, can be located at any position(s), and can vary in both intensity and texture. These aspects particularly aggravate the generation of an independent latent space capturing variability only (and do not contain information on e.g. muscle shape) which is an explicit prerequisite of~\cite{myAlmahairi18a,myHuang18a}. The proposed technique does not require a separate latent space and is thereby more generally applicable.



To conclude, we proposed a specific loss configuration facilitating training of cycle-GANs in the case of many-to-one mappings.
We showed, formally and empirically, that a slight modification of the cycle-consistency loss is capable of improving the performance without increasing the model complexity.
The experimental study on translating pathological into pseudo-healthy MR scans showed stable and significant improvement especially for the more difficult data set with a larger shift between the domains.
We are confident that the proposed method can improve performance also in other medical application scenarios with many-to-one mappings. 

\section{Compliance with Ethical Standards}
\label{sec:ethics}

This study was performed in line with the principles of the Declaration of Helsinki. Approval was granted by the Ethics Committee of RWTH Aachen University (2015/10/19, EK 278/15).

\section{Acknowledgments}
\label{sec:acknowledgments}

This work was supported by the county of Salzburg (FHS-2019-10-KIAMed) and the German Research Foundation (DFG: ME3737/3-1).


\bibliographystyle{IEEEbib}
\bibliography{my,eigene}

\end{document}